\documentclass[amsmath,aps,prd,showpacs,a4paper,10pt]{revtex4}

 \usepackage{epsf}
 \usepackage{graphicx}    

\begin{document}

 \newcommand{\be}[1]{\begin{equation}\label{#1}}
 \newcommand{\ee}{\end{equation}}
 \newcommand{\bea}{\begin{eqnarray}}
 \newcommand{\eea}{\end{eqnarray}}
 \def\disp{\displaystyle}

 \def\gsim{ \lower .75ex \hbox{$\sim$} \llap{\raise .27ex \hbox{$>$}} }
 \def\lsim{ \lower .75ex \hbox{$\sim$} \llap{\raise .27ex \hbox{$<$}} }

 \begin{titlepage}

 \begin{flushright}
 arXiv:0809.0057
 \end{flushright}

 \title{\Large \bf Cosmological Models and Latest Observational Data}

 \author{Hao~Wei\,}
 \email[\,email address:\ ]{haowei@bit.edu.cn}
 \affiliation{Department of Physics, Beijing Institute
 of Technology, Beijing 100081, China}

 \begin{abstract}\vspace{1cm}
 \centerline{\bf ABSTRACT}\vspace{2mm}
In this note, we consider the observational constraints on some
 cosmological models by using the 307 Union type Ia supernovae
 (SNIa), the 32 calibrated Gamma-Ray Bursts (GRBs) at $z>1.4$,
 the updated shift parameter $R$ from WMAP 5-year data (WMAP5),
 and the distance parameter $A$ of the measurement of the
 baryon acoustic oscillation (BAO) peak in the distribution of
 SDSS luminous red galaxies with the updated scalar spectral
 index $n_s$ from WMAP5. The tighter constraints obtained here
 update the ones obtained previously in the literature.
 \end{abstract}


 \maketitle

 \end{titlepage}

 \renewcommand{\baselinestretch}{1.6}



\section{Introduction}\label{sec1}
Recently, some observational data have been updated or have become
 available. In~\cite{r1,r2}, the Wilkinson Microwave Anisotropy
 Probe (WMAP) collaboration released their 5-year observational
 data (WMAP5). The data of Cosmic Microwave Background (CMB)
 anisotropy have been significantly improved. Also, in~\cite{r3,r4},
 the Supernova Cosmology Project (SCP) collaboration released their
 new dataset of type~Ia supernovae (SNIa), which was called the
 Union compilation. The Union compilation contains 414 SNIa
 and reduces to 307 SNIa after selection cuts. This 307 SNIa
 Union compilation is the currently largest SNIa dataset.

On the other hand, Gamma-Ray Bursts (GRBs) were proposed as
 a complementary probe to SNIa recently~\cite{r5,r6,r7,r8}.
 GRBs have been advocated as standard candles since several
 empirical GRB luminosity relations were proposed as distance
 indicators. However, there is the so-called circularity
 problem in the direct use of GRBs to probe
 cosmology~\cite{r5}. Recently, a new idea to calibrate GRBs in
 a completely {\em cosmology independent} manner has been
 proposed~\cite{r9,r10}, and the circularity problem can be
 solved. The main idea is that of the cosmic distance ladder.
 Similar to the case of calibrating SNIa as the secondary
 standard candles by using Cepheid variables, which are primary
 standard candles, we can also calibrate GRBs as standard
 candles with a large amount of SNIa. Following the calibration
 method proposed in~\cite{r10}, the distance moduli $\mu$ of 32
 calibrated GRBs at redshift $z>1.4$ are derived
 in~\cite{r11}. Now, one can use them to constrain cosmological
 models {\em without} circularity problem. See~\cite{r10,r11}
 for details. As argued in~\cite{r8,r32}, the observations at
 $z>1.7$ are fairly important to distinguish cosmological
 models and break the degeneracies between the parameters. In
 this note, we try to combine GRBs with the conventional
 datasets to constrain cosmological models. Although the number
 of GRBs is small and the systematic and statistical errors are
 very large so that their contribution to the constraints
 would be not so significant, this is still a beneficial
 exploration.

Here, we consider the observational constraints on some
 cosmological models by using the 307 Union SNIa compiled
 in~\cite{r3}, the 32 calibrated GRBs at $z>1.4$ compiled
 in Table~I of~\cite{r11}, the updated shift parameter $R$
 from WMAP5~\cite{r1}, and the distance parameter $A$ of
 the measurement of the baryon acoustic oscillation (BAO) peak
 in the distribution of SDSS luminous red galaxies~\cite{r12}
 with the updated scalar spectral index $n_s$ from
 WMAP5~\cite{r1}.

We perform a $\chi^2$ analysis to obtain the constraints on the
 parameters of cosmological models. The data points of the 307
 Union SNIa compiled in~\cite{r3} and the 32 calibrated GRBs at
 $z>1.4$ compiled in Table~I of~\cite{r11} are given in terms
 of the distance modulus $\mu_{obs}(z_i)$. On the other hand,
 the theoretical distance modulus is defined as
 \be{eq1}
 \mu_{th}(z_i)\equiv 5\log_{10}D_L(z_i)+\mu_0,
 \ee
 where $\mu_0\equiv 42.38-5\log_{10}h$ and $h$ is the Hubble
 constant $H_0$ in units of $100~{\rm km/s/Mpc}$, whereas
 \be{eq2}
 D_L(z)=(1+z)\int_0^z \frac{d\tilde{z}}{E(\tilde{z};{\bf p})},
 \ee
 in which $E\equiv H/H_0$ and $H$ is the Hubble parameter; ${\bf p}$
 denotes the model parameters. The $\chi^2$ from the 307 Union SNIa
 and the 32 calibrated GRBs at $z>1.4$ are given by
 \be{eq3}
 \chi^2_{\mu}({\bf p})=\sum\limits_{i}
 \frac{\left[\mu_{obs}(z_i)-\mu_{th}(z_i)\right]^2}{\sigma^2(z_i)},
 \ee
 where $\sigma$ is the corresponding $1\sigma$ error. The parameter
 $\mu_0$ is a nuisance parameter but it is independent of the data
 points. One can perform an uniform marginalization over $\mu_0$.
 However, there is an alternative way. Following~\cite{r13,r14}, the
 minimization with respect to $\mu_0$ can be made by expanding the
 $\chi^2_{\mu}$ of Eq.~(\ref{eq3}) with respect to $\mu_0$ as
 \be{eq4}
 \chi^2_{\mu}({\bf p})=\tilde{A}-2\mu_0\tilde{B}+\mu_0^2\tilde{C},
 \ee
 where
 $$\tilde{A}({\bf p})=\sum\limits_{i}\frac{\left[\mu_{obs}(z_i)
 -\mu_{th}(z_i;\mu_0=0,{\bf p})\right]^2}{\sigma_{\mu_{obs}}^2(z_i)}\,,$$
 $$\tilde{B}({\bf p})=\sum\limits_{i}\frac{\mu_{obs}(z_i)
 -\mu_{th}(z_i;\mu_0=0,{\bf p})}{\sigma_{\mu_{obs}}^2(z_i)}\,,
 ~~~~~~~~~~
 \tilde{C}=\sum\limits_{i}\frac{1}{\sigma_{\mu_{obs}}^2(z_i)}\,.$$
 Eq.~(\ref{eq4}) has a minimum for
 $\mu_0=\tilde{B}/\tilde{C}$ at
 \be{eq5}
 \tilde{\chi}^2_{\mu}({\bf p})=
 \tilde{A}({\bf p})-\frac{\tilde{B}({\bf p})^2}{\tilde{C}}.
 \ee
 Since $\chi^2_{\mu,\,min}=\tilde{\chi}^2_{\mu,\,min}$
 obviously, we can instead minimize $\tilde{\chi}^2_{\mu}$
 which is independent of $\mu_0$. Note that the above
 summations are over the 307 Union SNIa compiled in~\cite{r3}
 and the 32 calibrated GRBs at $z>1.4$ compiled in Table~I
 of~\cite{r11}. On the other hand, the shift parameter $R$ is
 defined by~\cite{r15,r16}
 \be{eq6}
 R\equiv\Omega_{m0}^{1/2}\int_0^{z_\ast}
 \frac{d\tilde{z}}{E(\tilde{z})},
 \ee
 where the redshift of recombination $z_\ast=1090$ which has
 been updated in~\cite{r1}, and $\Omega_{m0}$ is the present
 fractional energy density of pressureless matter. The shift
 parameter $R$ relates the angular diameter distance to the
 last scattering surface, the comoving size of the sound
 horizon at $z_\ast$ and the angular scale of the first
 acoustic peak in the CMB power spectrum of the temperature
 fluctuations~\cite{r15,r16}. The value of $R$ has been updated
 to $1.710\pm 0.019$ from WMAP5~\cite{r1}. The distance
 parameter $A$ is given by
 \be{eq7}
 A\equiv\Omega_{m0}^{1/2}E(z_b)^{-1/3}\left[\frac{1}{z_b}
 \int_0^{z_b}\frac{d\tilde{z}}{E(\tilde{z})}\right]^{2/3},
 \ee
 where $z_b=0.35$. In~\cite{r17}, the value of $A$ has been
 determined to be $0.469\,(n_s/0.98)^{-0.35}\pm 0.017$. Here
 the scalar spectral index $n_s$ is taken to be $0.960$ which
 has been updated from WMAP5~\cite{r1}. So, the total $\chi^2$
 is given by
 \be{eq8}
 \chi^2=\tilde{\chi}^2_{\mu}+\chi^2_{CMB}+\chi^2_{BAO},
 \ee
 where $\tilde{\chi}^2_{\mu}$ is given in Eq.~(\ref{eq5}),
 $\chi^2_{CMB}=(R-R_{obs})^2/\sigma_R^2$ and
 $\chi^2_{BAO}=(A-A_{obs})^2/\sigma_A^2$. The best-fit model
 parameters are determined by minimizing the total $\chi^2$.
 As in~\cite{r18}, the $68\%$ confidence level is determined by
 $\Delta\chi^2\equiv\chi^2-\chi^2_{min}\leq 1.0$, $2.3$ and
 $3.53$ for $n_p=1$, $2$ and $3$, respectively, where $n_p$ is
 the number of free model parameters. Similarly, the $95\%$
 confidence level is determined by
 $\Delta\chi^2\equiv\chi^2-\chi^2_{min}\leq 4.0$, $6.17$ and
 $8.02$ for $n_p=1$, $2$ and $3$, respectively.

In sections~\ref{sec2}, \ref{sec3} and~\ref{sec4}, we consider
 the joint constraints on single-parameter, two-parameter and
 three-parameter cosmological models respectively, by using
 the 307 Union SNIa compiled in~\cite{r3}, the 32 calibrated
 GRBs at $z>1.4$ compiled in Table~I of~\cite{r11}, the updated
 shift parameter $R$ from WMAP5~\cite{r1}, and the distance
 parameter $A$ of the measurement of BAO peak in the
 distribution of SDSS luminous red galaxies~\cite{r12} with the
 updated scalar spectral index $n_s$ from WMAP5~\cite{r1}. Note
 that we also present the constraints without GRBs for
 comparison. A brief summary is given in section~\ref{sec5}.


\section{Single-parameter models}\label{sec2}
In this section, we consider the constraints on three
 single-parameter models. They are the flat $\Lambda$CDM model,
 the flat DGP model and the new agegraphic dark energy (NADE)
 model.


\subsection{Flat $\Lambda$CDM model}\label{sec2a}
\begin{center}
 \begin{figure}[tbhp]
 \centering
 \includegraphics[width=0.49\textwidth]{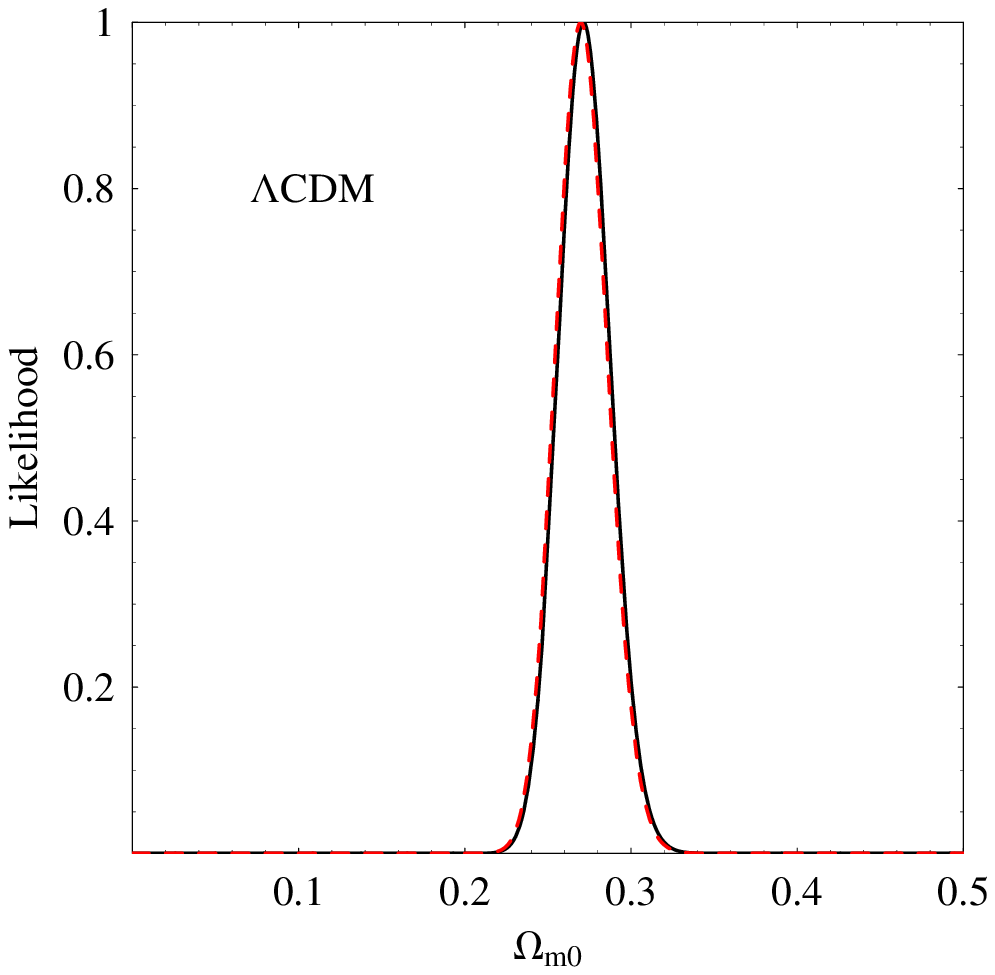}\hfill
 \includegraphics[width=0.49\textwidth]{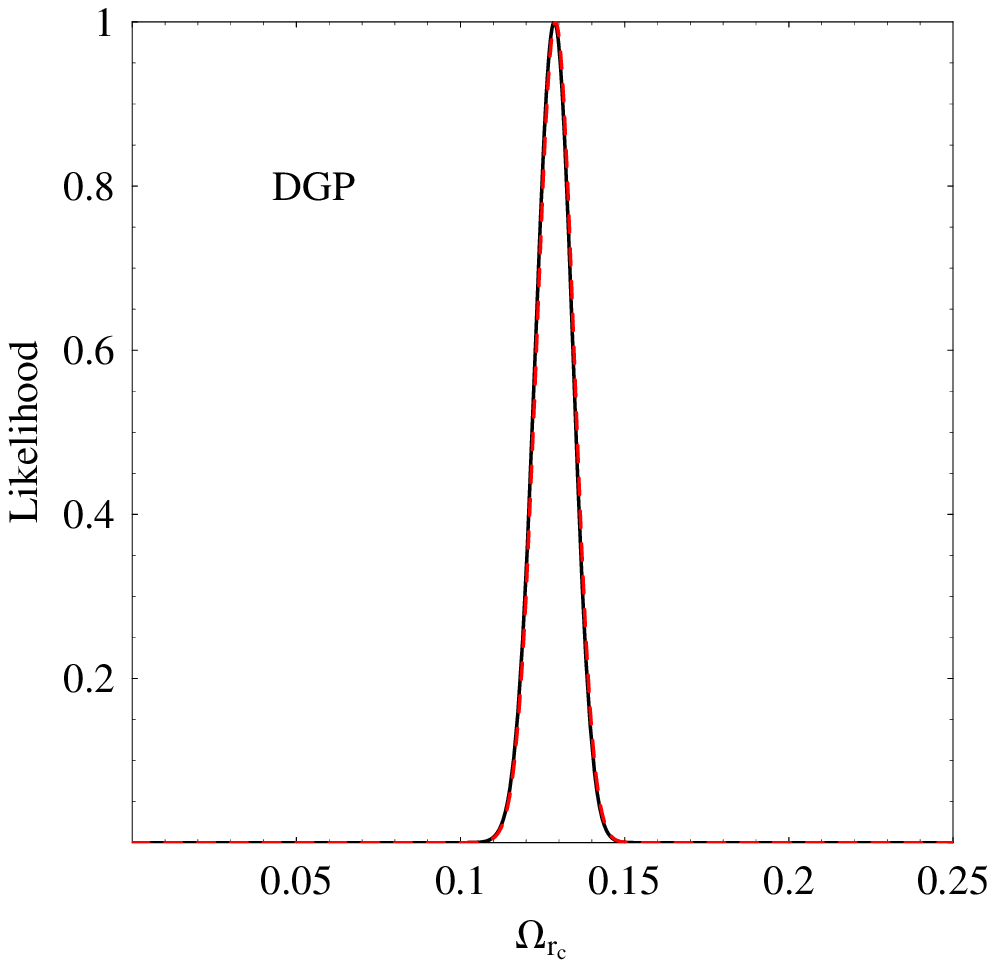}
 \caption{\label{fig1}
 The likelihood ${\cal L}\propto e^{-\chi^2/2}$ for the flat
 $\Lambda$CDM model and the flat DGP model. The results for the
 cases with and without GRBs are indicated by the black solid
 lines and the red dashed lines, respectively.}
 \end{figure}
 \end{center}

As is well known, for the spatially flat $\Lambda$CDM model,
 \be{eq9}
 E(z)=\sqrt{\Omega_{m0}(1+z)^3+(1-\Omega_{m0})}\,.
 \ee
 It is easy to obtain the total $\chi^2$ as a function of the single
 model parameter $\Omega_{m0}$. In Fig.~\ref{fig1}, we present the
 corresponding likelihood ${\cal L}\propto e^{-\chi^2/2}$. The best
 fit has $\chi^2_{min}=325.522$, whereas the best-fit parameter is
 \\ \vspace{-2.5mm}\\ 
 $\Omega_{m0}=0.2714^{+0.0159}_{-0.0152}$ (with $1\sigma$
 uncertainty) or $\Omega_{m0}=0.2714^{+0.0324}_{-0.0297}$ (with
 $2\sigma$ uncertainty).
 \\ \vspace{-2.5mm}\\ 
\indent For comparison, we also present the likelihood without GRBs
 in Fig.~\ref{fig1}, whereas the best-fit parameter
 \\ \vspace{-2.5mm}\\ 
 reads $\Omega_{m0}=0.2698^{+0.0159}_{-0.0152}$ (with $1\sigma$
 uncertainty) or $\Omega_{m0}=0.2698^{+0.0324}_{-0.0297}$ (with
 $2\sigma$ uncertainty).


\vspace{-10mm}  


\subsection{Flat DGP model}\label{sec2b}
One of the leading modified gravity models is the so-called
 Dvali-Gabadadze-Porrati~(DGP) braneworld model~\cite{r19,r20},
 which entails altering the Einstein-Hilbert action by a term
 arising from large extra dimensions. For a list of references
 on the DGP model, see e.g.~\cite{r21,r22} and references
 therein.

As is well known, for the spatially flat DGP model (here we
 only consider the self-accelerating branch), $E(z)$ is given
 by~\cite{r20,r21,r22}
 \be{eq10}
 E(z)=\sqrt{\Omega_{m0}(1+z)^3+\Omega_{r_c}}+
 \sqrt{\Omega_{r_c}}\,,
 \ee
 where $\Omega_{r_c}$ is constant. $E(z=0)=1$
 requires
 \be{eq11}
 \Omega_{m0}=1-2\sqrt{\Omega_{r_c}}\,.
 \ee
 Therefore, the flat DGP model has only one independent model
 parameter. Notice that $0\leq\Omega_{r_c}\leq 1/4$ is required
 by $0\leq\Omega_{m0}\leq 1$.

It is easy to obtain the total $\chi^2$ as a function of the single
 model parameter $\Omega_{r_c}$. Also in Fig.~\ref{fig1}, we plot the
 corresponding likelihood ${\cal L}\propto e^{-\chi^2/2}$. The best
 fit has $\chi^2_{min}=345.56$, whereas the best-fit parameter
 \\ \vspace{-2.5mm}\\ 
 is $\Omega_{r_c}=0.1286^{+0.0056}_{-0.0057}$ (with $1\sigma$
 uncertainty) or $\Omega_{r_c}=0.1286^{+0.0111}_{-0.0116}$ (with
 $2\sigma$ uncertainty). From Eq.~(\ref{eq11}),
 \\ \vspace{-2.5mm}\\ 
 $\Omega_{m0}$ can be derived correspondingly.
 \\ \vspace{-2.5mm}\\ 
\indent For comparison, we also present the likelihood without GRBs
 in Fig.~\ref{fig1}, whereas the best-fit parameter
 \\ \vspace{-2.5mm}\\ 
 reads $\Omega_{r_c}=0.1289^{+0.0056}_{-0.0057}$ (with $1\sigma$
 uncertainty) or $\Omega_{r_c}=0.1289^{+0.0111}_{-0.0116}$ (with
 $2\sigma$ uncertainty).


\subsection{New agegraphic dark energy model}\label{sec2c}
In~\cite{r23,r24}, the so-called ``new agegraphic dark energy''
 (NADE) model has been proposed recently, based on the
 K\'{a}rolyh\'{a}zy uncertainty relation which arises from
 quantum mechanics together with general relativity. In this
 model, the energy density of NADE is given by~\cite{r23,r24}
 \be{eq12}
 \rho_q=\frac{3n^2m_p^2}{\eta^2}\,,
 \ee
 where $m_p$ is the reduced Planck mass; $n$ is a constant of
 order unity; $\eta$ is the conformal time
 \be{eq13}
 \eta\equiv\int\frac{dt}{a}=\int\frac{da}{a^2H}\,,
 \ee
 in which $a=(1+z)^{-1}$ is the scale factor. Obviously,
 $\dot{\eta}=1/a$, where a dot denotes the derivative with
 respect to cosmic time $t$. The corresponding fractional
 energy density of NADE reads
 \be{eq14}
 \Omega_q\equiv\frac{\rho_q}{3m_p^2H^2}=\frac{n^2}{H^2\eta^2}.
 \ee
 From the Friedmann equation
 $H^2=\left(\rho_m+\rho_q\right)/\left(3m_p^2\right)$,
 the energy conservation equation $\dot{\rho}_m+3H\rho_m=0$,
 and Eqs.~(\ref{eq12})---(\ref{eq14}), we find that the
 equation of motion for $\Omega_q$ is given by~\cite{r23,r24}
 \be{eq15}
 \frac{d\Omega_q}{dz}=-\Omega_q\left(1-\Omega_q\right)
 \left[3(1+z)^{-1}-\frac{2}{n}\sqrt{\Omega_q}\right].
 \ee
 From the energy conservation equation
 $\dot{\rho}_q+3H(\rho_q+p_q)=0$, and
 Eqs.~(\ref{eq12})---(\ref{eq14}), it is easy to find that
 the equation-of-state parameter~(EoS) of NADE is given
 by~\cite{r23,r24}
 \be{eq16}
 w_q\equiv\frac{p_q}{\rho_q}=
 -1+\frac{2}{3n}\frac{\sqrt{\Omega_q}}{a}.
 \ee
 When $a\to\infty$, $\Omega_q\to 1$, thus $w_q\to -1$ in the
 late time. When $a\to 0$, $\Omega_q\to 0$, so $0/0$ appears in
 $w_q$ and hence we cannot obtain $w_q$ from Eq.~(\ref{eq16})
 directly. Let us consider the matter-dominated epoch,
 $H^2\propto\rho_m\propto a^{-3}$. Thus,
 $a^{1/2}da\propto dt=ad\eta$. Therefore, $\eta\propto a^{1/2}$.
 From Eq.~(\ref{eq12}), $\rho_q\propto a^{-1}$.  From the energy
 conservation equation $\dot{\rho}_q+3H\rho_q(1+w_q)=0$, we
 obtain $w_q=-2/3$ in the matter-dominated epoch. Since
 $\rho_m\propto a^{-3}$ and $\rho_q\propto a^{-1}$, it is
 expected that $\Omega_q\propto a^2$. Comparing $w_q=-2/3$ with
 Eq.~(\ref{eq16}), we find that $\Omega_q=n^2a^2/4$ in the
 matter-dominated epoch as expected. For $a\ll 1$, provided
 that $n$ is of order unity, $\Omega_q\ll 1$ naturally follows.
 There are many interesting features in the NADE model and we
 refer to the original papers~\cite{r23,r24} for more details.

At first glance, one might consider that NADE is
 a two-parameter model. However, as shown in~\cite{r23}, NADE
 is a {\em single-parameter} model in practice, thanks to its
 special analytic feature $\Omega_q=n^2a^2/4=n^2(1+z)^{-2}/4$
 in the matter-dominated epoch, as mentioned above. If $n$ is
 given, we can obtain $\Omega_q(z)$ from Eq.~(\ref{eq15}) with
 the initial condition $\Omega_q(z_{ini})=n^2(1+z_{ini})^{-2}/4$ at
 any $z_{ini}$ which is deep enough into the matter-dominated
 epoch (we choose $z_{ini}=2000$ as in~\cite{r23}), instead of
 $\Omega_q(z=0)=\Omega_{q0}=1-\Omega_{m0}$ at $z=0$. Then, all
 other physical quantities, such as $\Omega_m(z)=1-\Omega_q(z)$
 and $w_q(z)$ in Eq.~(\ref{eq16}), can be obtained correspondingly.
 So, $\Omega_{m0}=\Omega_m(z=0)$, $\Omega_{q0}=\Omega_q(z=0)$ and
 $w_{q0}=w_q(z=0)$ are {\em not} independent model parameters. The
 only model parameter is $n$. Therefore, the NADE model is a
 {\em single-parameter} model in practice. To our knowledge, it
 is the third single-parameter cosmological model besides the
 flat $\Lambda$CDM model and the flat DGP model.

\begin{center}
 \begin{figure}[htbp]
 \centering
 \includegraphics[width=0.5\textwidth]{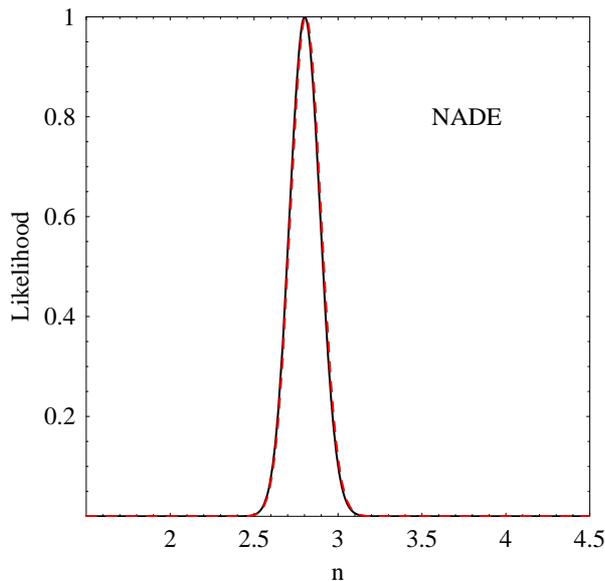}
 \caption{\label{fig2}
 The likelihood ${\cal L}\propto e^{-\chi^2/2}$ for the NADE
 model. The results for the cases with and without GRBs are
 indicated by the black solid line and the red dashed line,
 respectively.}
 \end{figure}
 \end{center}

From the Friedmann equation
 $H^2=\left(\rho_m+\rho_q\right)/\left(3m_p^2\right)$, we have
 \be{eq17}
 E(z)=\left[\frac{\Omega_{m0}(1+z)^3}{1-\Omega_q(z)}\right]^{1/2}.
 \ee
 If the single model parameter $n$ is given, we can obtain
 $\Omega_q(z)$ from Eq.~(\ref{eq15}). Thus, we get
 $\Omega_{m0}=1-\Omega_q(z=0)$. So, $E(z)$ is at hand.
 Therefore, we can find the corresponding total $\chi^2$ in
 Eq.~(\ref{eq8}). In Fig.~\ref{fig2}, we plot the
 corresponding likelihood ${\cal L}\propto e^{-\chi^2/2}$ as a
 function of $n$. The best fit has $\chi^2_{min}=336.061$,
 whereas the best-fit parameter is $n=2.802$. We present
 the best-fit value of $n$ and the corresponding derived
 $\Omega_{m0}$, $\Omega_{q0}$ and $w_{q0}$ with $1\sigma$ and
 $2\sigma$ uncertainties in Table~\ref{tab1}. Obviously, these
 constraints on the NADE model are tighter than the ones
 obtained in~\cite{r23}.

For the case without GRBs, the best-fit parameter is $n=2.808$.
 For comparison, we also present the best-fit value of $n$ and
 the corresponding derived $\Omega_{m0}$, $\Omega_{q0}$ and
 $w_{q0}$ with $1\sigma$ and $2\sigma$ uncertainties in
 Table~\ref{tab1}.

 \begin{table}[tbhp]
 \begin{center}
 \begin{tabular}{c|c|c|c|c} \hline\hline
 Uncertainty \ \ & $n$ & $\Omega_{m0}$ & $\Omega_{q0}$ & $w_{q0}$
 \\ \hline
 $1\sigma$ (with GRBs)\ \ & \ \ $2.802^{+0.092}_{-0.090}$\ \
 & \ \ $0.279^{+0.016}_{-0.015}$\ \
 & \ \ $0.721^{+0.015}_{-0.016}$\ \
 & \ \ $-0.798^{+0.004}_{-0.004}$\ \ \\
 $2\sigma$ (with GRBs)\ \ & \ \ $2.802^{+0.185}_{-0.179}$\ \
 & \ \ $0.279^{+0.033}_{-0.030}$\ \
 & \ \ $0.721^{+0.030}_{-0.033}$\ \
 & \ \ $-0.798^{+0.009}_{-0.009}$\ \ \\ \hline\hline
 $1\sigma$ (without GRBs)\ \ & \ \ $2.808^{+0.092}_{-0.090}$\ \
 & \ \ $0.278^{+0.016}_{-0.015}$\ \
 & \ \ $0.722^{+0.015}_{-0.016}$\ \
 & \ \ $-0.798^{+0.004}_{-0.004}$\ \ \\
 $2\sigma$ (without GRBs)\ \ & \ \ $2.808^{+0.186}_{-0.179}$\ \
 & \ \ $0.278^{+0.033}_{-0.030}$\ \
 & \ \ $0.722^{+0.030}_{-0.033}$\ \
 & \ \ $-0.798^{+0.009}_{-0.009}$\ \ \\ \hline\hline
 \end{tabular}
 \end{center}
 \caption{\label{tab1} The best-fit value of $n$ and the
 corresponding derived $\Omega_{m0}$, $\Omega_{q0}$ and
 $w_{q0}$ with $1\sigma$ and $2\sigma$ uncertainties for the
 NADE model. See text for details.}
 \end{table}


\vspace{-12mm}  


\section{Two-parameter model}\label{sec3}
\begin{center}
 \begin{figure}[htbp]
 \centering
 \includegraphics[width=0.5\textwidth]{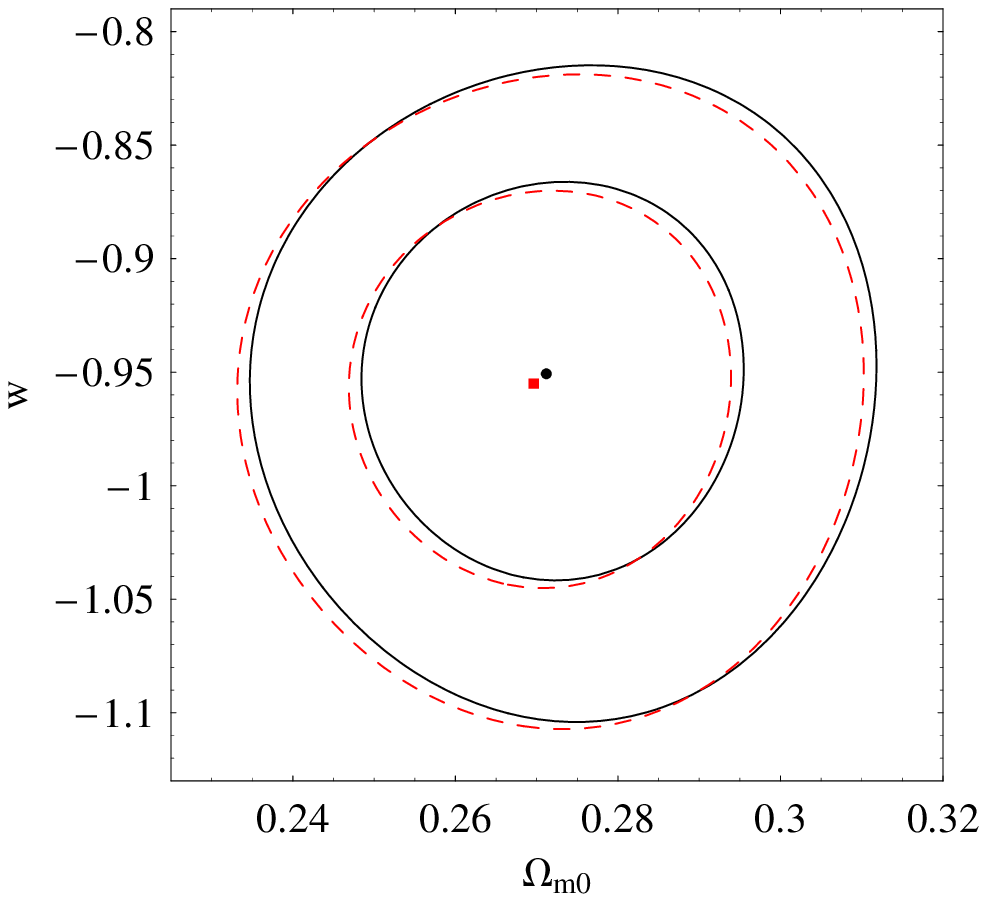}
 \caption{\label{fig3}
 The $68\%$ and $95\%$ confidence level contours in the
 $\Omega_{m0}-w$ parameter space for the XCDM model. The
 results for the cases with and without GRBs are indicated
 by the black solid lines and the red dashed lines,
 respectively. The best-fit parameters for the cases with
 and without GRBs are also indicated by a black solid point
 and a red box, respectively.}
 \end{figure}
 \end{center}

Here, we consider the XCDM model which is a two-parameter
 model. In the spatially flat universe which contains
 pressureless matter and dark energy whose EoS is a
 constant $w$, the corresponding $E(z)$ is given by
 \be{eq18}
 E(z)=\sqrt{\Omega_{m0}(1+z)^3+(1-\Omega_{m0})(1+z)^{3(1+w)}}\,.
 \ee
 By minimizing the corresponding total $\chi^2$ in Eq.~(\ref{eq8}),
 we find the best-fit parameters $\Omega_{m0}=0.271$ and
 $w=-0.951$, while $\chi^2_{min}=324.821$. In Fig.~\ref{fig3},
 we present the corresponding $68\%$ and $95\%$ confidence level
 contours in the $\Omega_{m0}-w$ parameter space for the XCDM model.
 For comparison, the best-fit parameters are $\Omega_{m0}=0.270$ and
 $w=-0.954$ for the case without GRBs, whereas the corresponding
 confidence level contours are also presented in Fig.~\ref{fig3}.


\vspace{-12mm}  


\section{Three-parameter model}\label{sec4}
\begin{center}
 \begin{figure}[tbp]
 \centering
 \includegraphics[width=0.5\textwidth]{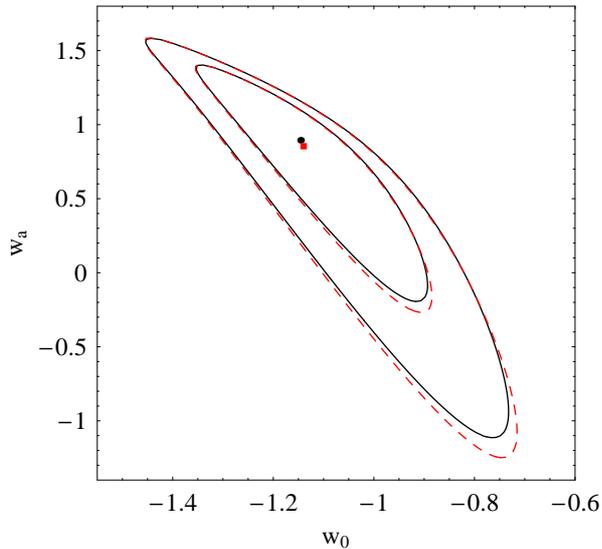}
 \caption{\label{fig4}
 The $68\%$ and $95\%$ confidence level contours in the
 $w_0-w_a$ plane for the CPL model. The results for the cases
 with and without GRBs are indicated by the black solid lines
 and the red dashed lines, respectively. The best-fit parameters
 for the cases with and without GRBs are also indicated by a
 black solid point and a red box, respectively.}
 \end{figure}
 \end{center}

\begin{center}
 \begin{figure}[htbp]
 \centering
 \includegraphics[width=1.0\textwidth]{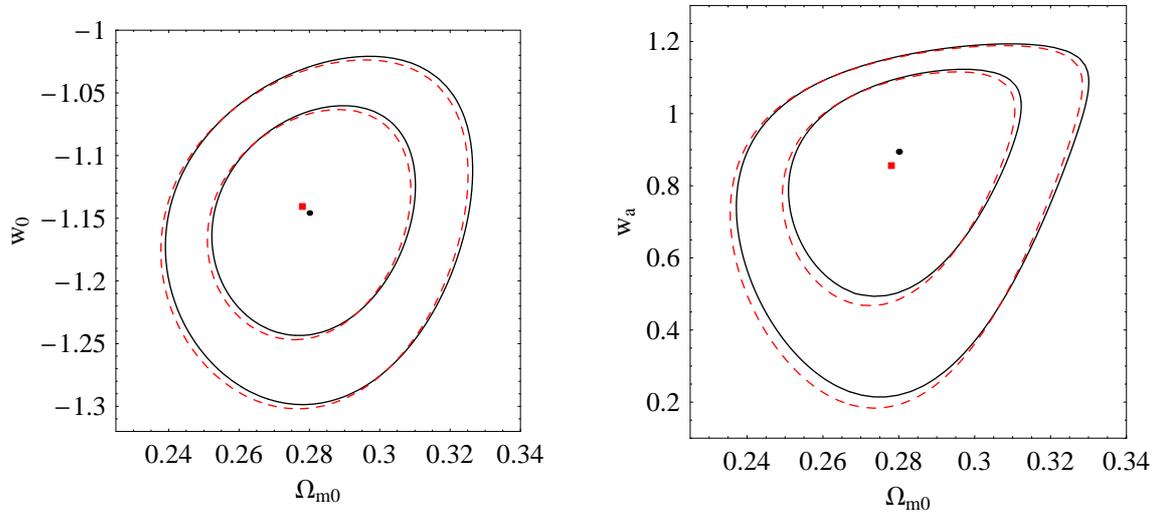}
 \caption{\label{fig5}
 The same as in Fig.~\ref{fig4}, except for the $\Omega_{m0}-w_0$
 plane and the $\Omega_{m0}-w_a$ plane.}
 \end{figure}
 \end{center}

Now, we consider the familiar Chevallier-Polarski-Linder
 (CPL) model~\cite{r25,r26}, in which the EoS of dark energy is
 parameterized as
 \be{eq19}
 w_{de}=w_0+w_a(1-a)=w_0+w_a\frac{z}{1+z},
 \ee
 where $w_0$ and $w_a$ are constants. As is well known, the
 corresponding $E(z)$ is given by~\cite{r18,r27,r28}
 \be{eq20}
 E(z)=\left[\Omega_{m0}(1+z)^3
 +\left(1-\Omega_{m0}\right)(1+z)^{3(1+w_0+w_a)}
 \exp\left(-\frac{3w_a z}{1+z}\right)\right]^{1/2}.
 \ee
 There are 3 independent parameters in this model. By minimizing the
 corresponding total $\chi^2$ in Eq.~(\ref{eq8}), we find the
 best-fit parameters $\Omega_{m0}=0.280$, $w_0=-1.146$
 and $w_a=0.894$, while $\chi^2_{min}=322.475$. In Fig.~\ref{fig4},
 we present the corresponding $68\%$ and $95\%$ confidence level
 contours in the $w_0-w_a$ plane for the CPL model. Also,
 the $68\%$ and $95\%$ confidence level contours in the
 $\Omega_{m0}-w_0$ plane and the $\Omega_{m0}-w_a$ plane for
 the CPL model are shown in Fig.~\ref{fig5}. It is easy to see
 that these constraints on the CPL model are much tighter than
 the ones obtained in~\cite{r29}.

For comparison, the best-fit parameters are $\Omega_{m0}=0.278$,
 $w_0=-1.140$ and $w_a=0.859$ for the case without GRBs, whereas
 the corresponding confidence level contours are also presented
 in Figs.~\ref{fig4} and~\ref{fig5}.


\section{Summary}\label{sec5}
In this note, we consider the observational constraints on some
 cosmological models by using the 307 Union SNIa compiled
 in~\cite{r3}, the 32 calibrated GRBs at $z>1.4$ compiled in
 Table~I of~\cite{r11}, the updated shift parameter $R$ from
 WMAP5~\cite{r1}, and the distance parameter $A$ of the
 measurement of the baryon acoustic oscillation (BAO) peak in
 the distribution of SDSS luminous red galaxies~\cite{r12} with
 the updated scalar spectral index $n_s$ from WMAP5~\cite{r1}.
 The tighter constraints obtained here update the ones obtained
 previously in the literature
 (e.g.~\cite{r13,r14,r18,r23,r28,r29,r30,r31}).

It is worth noting that GRBs are potential tools which
 might be powerful to probe the cosmic expansion history up to
 $z>6$ or even higher redshift. Of course, due to the large
 scatter and the lack of a large amount of well observed GRBs,
 there is a long way to go in using GRBs extensively and
 reliably to probe cosmology. The cosmology independent calibration
 method of GRBs proposed in~\cite{r10} is an important step towards
 this end. The works of this note and~\cite{r11} are beneficial
 explorations. Along with the accumulation of well observed
 GRBs with much smaller errors, we believe that a bright future
 of GRB cosmology is awaiting us. Combining the calibrated GRBs with
 other probes such as SNIa, CMB, large-scale structure and weak
 lensing, we can learn more about the mysterious dark energy.


\vspace{-10mm}  


\section*{ACKNOWLEDGEMENTS}
We thank the anonymous referee for quite useful comments and
 suggestions, which help us to improve this work. We are
 grateful to Prof. Shuang~Nan~Zhang and Prof. Rong-Gen~Cai
 for helpful discussions. We also thank Minzi~Feng, as well as
 Nan~Liang, Yuan~Liu, Wei-Ke~Xiao, Pu-Xun~Wu, Rong-Jia~Yang,
 Jian~Wang, and Bin~Shao, Yu~Tian, Zhao-Tan~Jiang, Feng~Wang,
 Jian~Zou, Zhi~Wang, Xiao-Ping~Jia, for kind help and
 discussions. This work was supported by the Excellent Young
 Scholars Research Fund of Beijing Institute of Technology.


\vspace{-10mm}  

 \renewcommand{\baselinestretch}{1.1}


\end{document}